\documentclass[twocolumn, 10pt, secnumarabic, amssymb, nobibnotes]{revtex4-1}

\usepackage{amsmath,mathrsfs,latexsym,amssymb,graphicx,xcolor,float,setspace,
graphicx,subfigure}

\usepackage[colorlinks=blue,linkcolor=blue]{hyperref}%

\begin{document}

\title{\Large{Tailoring the slow light behavior in terahertz metasurfaces}}

\author{Manukumara Manjappa}
\affiliation{\footnotesize{Center for Disruptive Photonic Technologies, Division of Physics and Applied Physics, School of Physical and Mathematical Sciences, Nanyang Technological University, 21 Nanyang Link, Singapore 637371.}}
\author{Sher-Yi Chiam}
\affiliation{\footnotesize{Department of Physics, National University of Singapore, Science Drive 3, Singapore 117542, Singapore.}}
\affiliation{\footnotesize{NUS High School of Math and Science, 20 Clementi Avenue 1, Singapore 129957, Singapore.}}%
\author{Longqing Cong}
\affiliation{\footnotesize{Center for Disruptive Photonic Technologies, Division of Physics and Applied Physics, School of Physical and Mathematical Sciences, Nanyang Technological University, 21 Nanyang Link, Singapore 637371.}}
\author{Andrew A. Bettiol}
\affiliation{\footnotesize{Department of Physics, National University of Singapore, Science Drive 3, Singapore 117542, Singapore.}}
\author{Weili Zhang}
\affiliation{\footnotesize{School of Electrical Engineering and Computer Science, Oklahoma State University, Stillwater, Oklahoma 87074, USA.}}
\author{Ranjan Singh}
\email{ranjans@ntu.edu.sg}
\affiliation{\footnotesize{Center for Disruptive Photonic Technologies, Division of Physics and Applied Physics, School of Physical and Mathematical Sciences, Nanyang Technological University, 21 Nanyang Link, Singapore 637371.}}

\begin{abstract}
We experimentally study the effect of near field coupling on the transmission of light in terahertz metasurfaces, possessing slightly distinctive SRR resonances. Our results show that the interplay between the strengths of electric and magnetic dipoles, modulates the amplitude of resulting electromagnetically induced transmission, probed under different types of asymmetries in the coupled system. We employ a two-particle model to theoretically study the influence of the near field coupling between bright and quasi-dark modes on the transmission properties of the coupled system and we find an excellent agreement with our observed results. Adding to the enhanced transmission characteristics, our results provide a deeper insight into the metamaterial analogues of atomic electromagnetically induced transparency and offer an approach to engineer slow light devices, broadband filters and attenuators at terahertz frequencies.     
\end{abstract}

\maketitle
Light-matter interaction has been a subject of intense research over past several decades, since it allows to probe the resonance and the off-resonance properties of the materials over large part of the electromagnetic spectrum. Until late twentieth century, light-matter interaction in the terahertz part of the electromagnetic spectrum was the least explored. With the advent of metamaterials\cite{1,2,3}, which exhibit structure dependent resonance properties, have become excellent candidates for probing such resonant and off-resonant interactions at terahertz frequencies. Metamaterials are composed of periodic array of sub wavelength sized meta-atoms, which exhibit strong near-field coupling that can carry the interaction energy over to the far field regimes. Superlens\cite{4,5}, hybridization\cite{6,7,8,9}, Fano-coupling\cite{10,11,12} and the classical analogue of electromagnetically induced transparency (EIT)\cite{13,14,15,16,17} have been studied and demonstrated using the near field coupling within the metamaterials. Recently, there have been a enormous interest in the near-field coupling in terahertz metamaterials, which show EIT like transmission\cite{18,19,20} and ultra high Q Fano resonances\cite{21,22}, which find significant applications in the terahertz sensing\cite{23,24} and broadband communication technologies\cite{25}.     

Electromagnetically induced transparency is a quantum interference effect, which was first observed\cite{26} in a three level atomic system, owing to the destructive interference between the possible excitation pathways. Later its analogue was extended to the classical systems\cite{27}, since then EIT effects have been observed in various classical systems, including metamaterials\cite{13,14,15,16,17,18,19}, photonic crystals\cite{28}, micro ring resonators\cite{29,30} and all dielectric metasurfaces\cite{31}.  There have been a few reports on tailoring the classical analogue of EIT using metamaterials at microwave\cite{32,33,34}, terahertz\cite{35,36,37} and optical frequencies\cite{13,38}, either by tuning the near field coupling or by changing the material properties. Manipulation of EIT  in classical systems will allow us to precisely tailor the group velocity\cite{19,38} and the delay bandwidth product\cite{32} of the transmitted pulse. Moreover, it provides a clear picture of the coupling mechanisms in the classical analogue of EIT, that can help us in drawing the closest analogy between the classical and the quantum systems.     

 \begin{figure}[t]
\begin{center}
  \includegraphics[width=8cm]{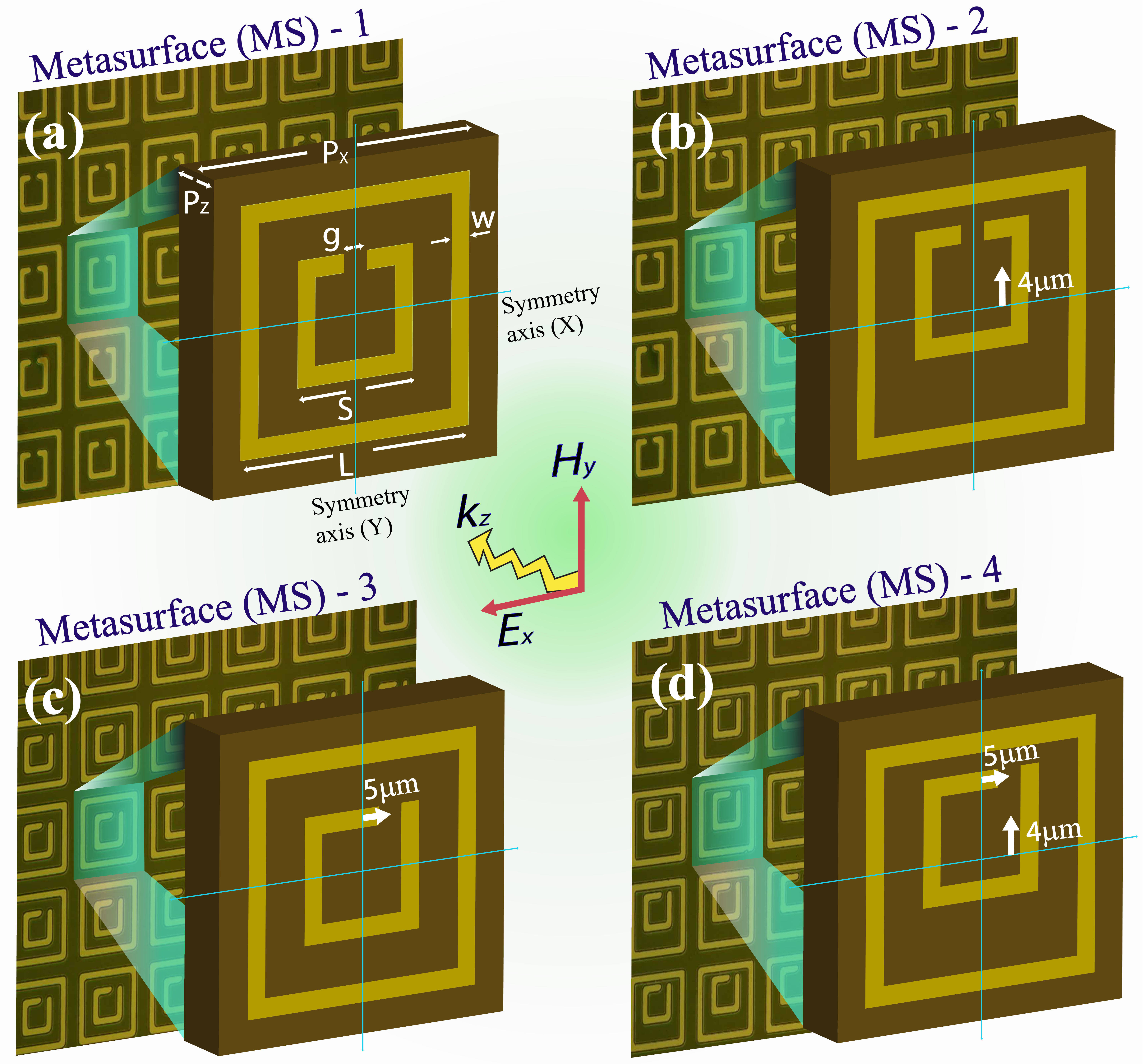}
\caption{\footnotesize{(color online). Metasurfaces showing {\bf{(a)}} symmetry (MS-1), {\bf{(b)}} SRR-position asymmetry (MS-2), {\bf{(c)}} SRR-gap asymmetry (MS-3) and {\bf{(d)}} SRR-position and SRR-gap asymmetry (MS-4), are graphically displayed. In the case of {\bf{(b)}} and {\bf{(d)}} SRR ring is displaced upwards by 4$\mu$m from the symmetric axis (x), whereas in {\bf{(c)}} and {\bf{(d)}} SRR gap is displaced sidewards by 5$\mu$m from the symmetric axis (Y). All four metasurface samples have same material dimension: L, 40$\mu$m; S, 20$\mu$m; g, 4$\mu$m; w, 3$\mu$m; Periodicity ($P_{x}$) of the unit cell is 50$\mu$m with substrate thickness ($P_{z}$) of 640$\mu$m.}}
\end{center}
\end{figure}

In this letter, we experimentally demonstrate the enhancement and suppression in transmission of the fields at terahertz frequencies, by manipulating the near field coupling between the radiative dipole ring and the sub-radiant quasi-dark split ring resonator (SRR) ring in metasurfaces under different type of asymmetries of the metamolecule. Systems with enhanced transmission show a considerable increase in the delay bandwidth product (DBP) at the transmission peak. We quantitatively interpret underlying phenomenon by using the two-particle model, which shows an excellent agreement with the observed experimental results and is discussed in detail in the following sections. 

Metamaterial unit cells (shown in Fig.1), consist of a metallic split ring resonator (SRR) surrounded by a concentric metallic closed square ring resonator (CRR), both having a thickness of 200nm. Samples were fabricated using photolithography technique, where 200nm thin layer of aluminium is deposited on 640$\mu m$ thick n-type silicon substrate ($\epsilon =$ 11.68). Structural symmetry of the metamolecule unit cells were broken to study the impact of the near field coupling in the asymmetric metasurface array. Metasurfaces, MS-3 and MS-4, are the broken SRR-gap symmetry structures of MS-1 and MS-2 respectively (categorised as SRR-gap asymmetry), whereas, MS-2 and MS-4 are the broken SRR-position symmetries of MS-1 and MS-3 respectively (categorized as SRR-position asymmetry). For the metasurfaces (MS-3 and MS-4) with SRR-gap asymmetry, the capacitive split gap in the SRRs of MS-1 and MS-2 is displaced horizontally (along the {\emph{x}}-axis) by 5$\mu$m from the {\emph{y}}-symmetry axis, whereas for the metasurfaces with SRR-position asymmetry (MS-2 and MS-4), position of the inner SRR ring in MS-1 and MS-3 is displaced upwards (along the {\emph{y}}-axis) by 4$\mu$m from the {\emph{x}}-symmetry axis (see Fig.1). The design of each metamolecule was chosen such that the fundamental resonance frequencies of their constituent resonators, exhibiting highly contrasting resonance linewidths, fall at the same frequency, which is essential to realize the EIT like behavior in classical systems. Fig.2(a) depicts the contrasting resonance linewidths for the CRR ring and the SRRs. The measured {\emph{Q}}-factor of CRR is 1.2 which is an order of magnitude lower than that of the {\emph{Q}}-factor of inner SRR. For the symmetric SRR, the {\emph{Q}}-factor is 10.6, whereas the asymmetric SRR (SRR with displaced gap) has a {\emph{Q}}-factor of 11.3. Here, we would like to stress that both the resonators, CRR and the SRR interact with the incoming electric field ($E_{x}$) but with very different coupling strengths, which is determined by the {\emph{Q}}-factors of their fundamental resonances. The SRR ring with sharp {\emph{LC}} resonance (higher quality factor) is termed as `quasi-dark mode', because of its weak coupling to the incoming light field ($E_{x}$), where as the CRR ring with broad dipolar resonance (lower quality factor) behaves like a `bright mode' that couples strongly to the incoming light field ($E_{x}$). 

\begin{figure}[t]
\begin{center}
  \includegraphics[width=8cm]{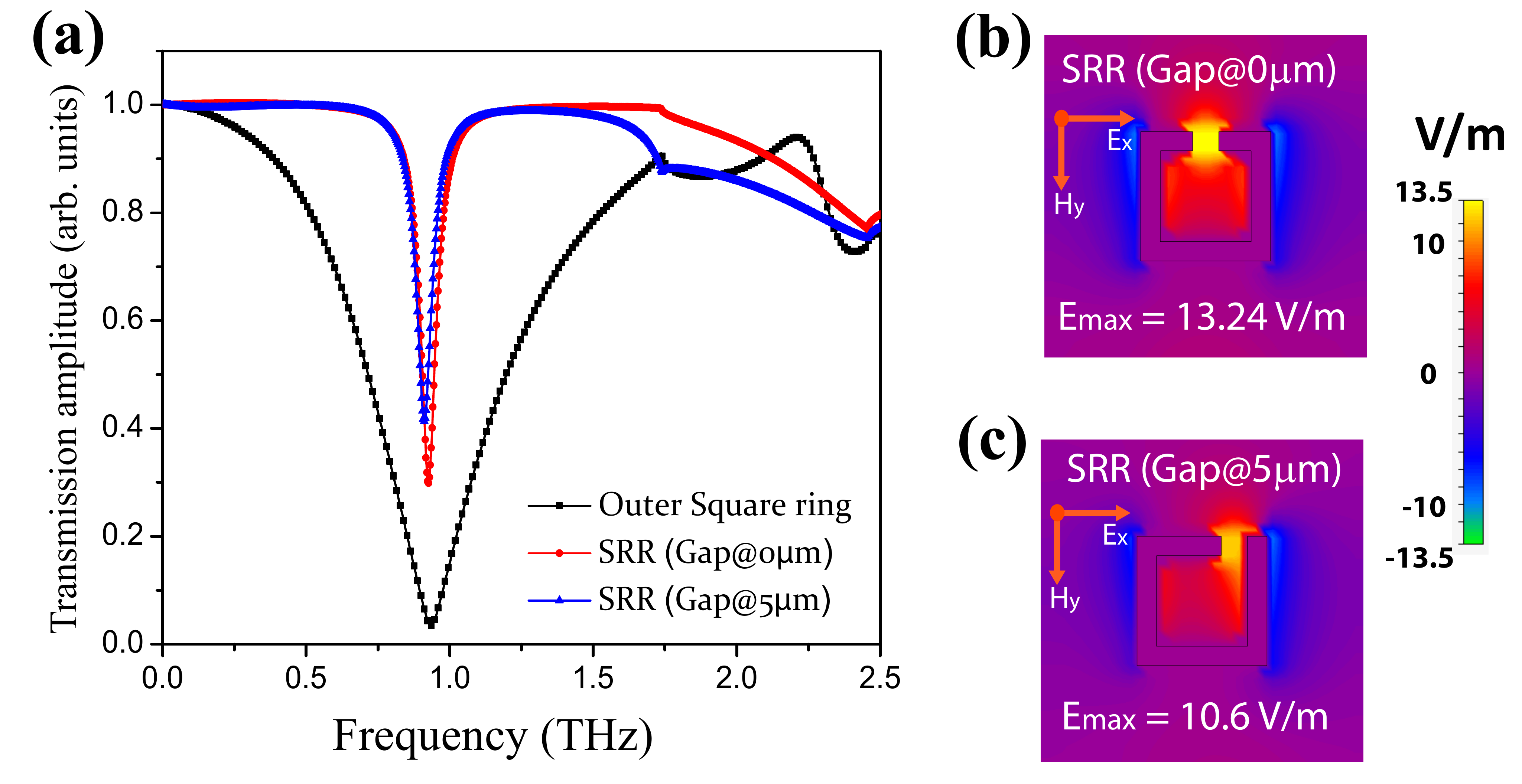}
\caption{\footnotesize{(color online). {\bf{(a)}} Transmission vs frequency graph displaying the individual resonance dips for the outer closed resonator ring (CRR) (black), the inner symmetric SRR (red) and the inner asymmetric SRR (blue) rings. Electric field strengths for the symmetric {\bf{(b)}} and the asymmetric {\bf{(c)}} SRR rings are also shown. }}
\end{center}
\end{figure}

Figures 3(a)-(d) show the sharp transmission spectra for each of the metamaterial samples. Spectra were recorded using 8f confocal terahertz time domain spectroscopy (THz-TDS) for the incident electric field ($E_{x}$), polarized along the gap ({\emph{x}}-axis) of the SRR ring. Recorded transmission time domain signals were converted to frequency domain data using FFT and normalized to the transmission of the bare silicon substrate (as reference). Such narrow transmission peaks result from the destructive interference between the classical resonators (CRR and SRR) due to their strong near field coupling. Corresponding numerical simulations were carried out using the commercially available CST MICROWAVE STUDIO Maxwell equation solver and the data matched well with our measured results (insets of Fig.3).  

\begin{figure}[t]
\begin{center}
  \includegraphics[width=9cm]{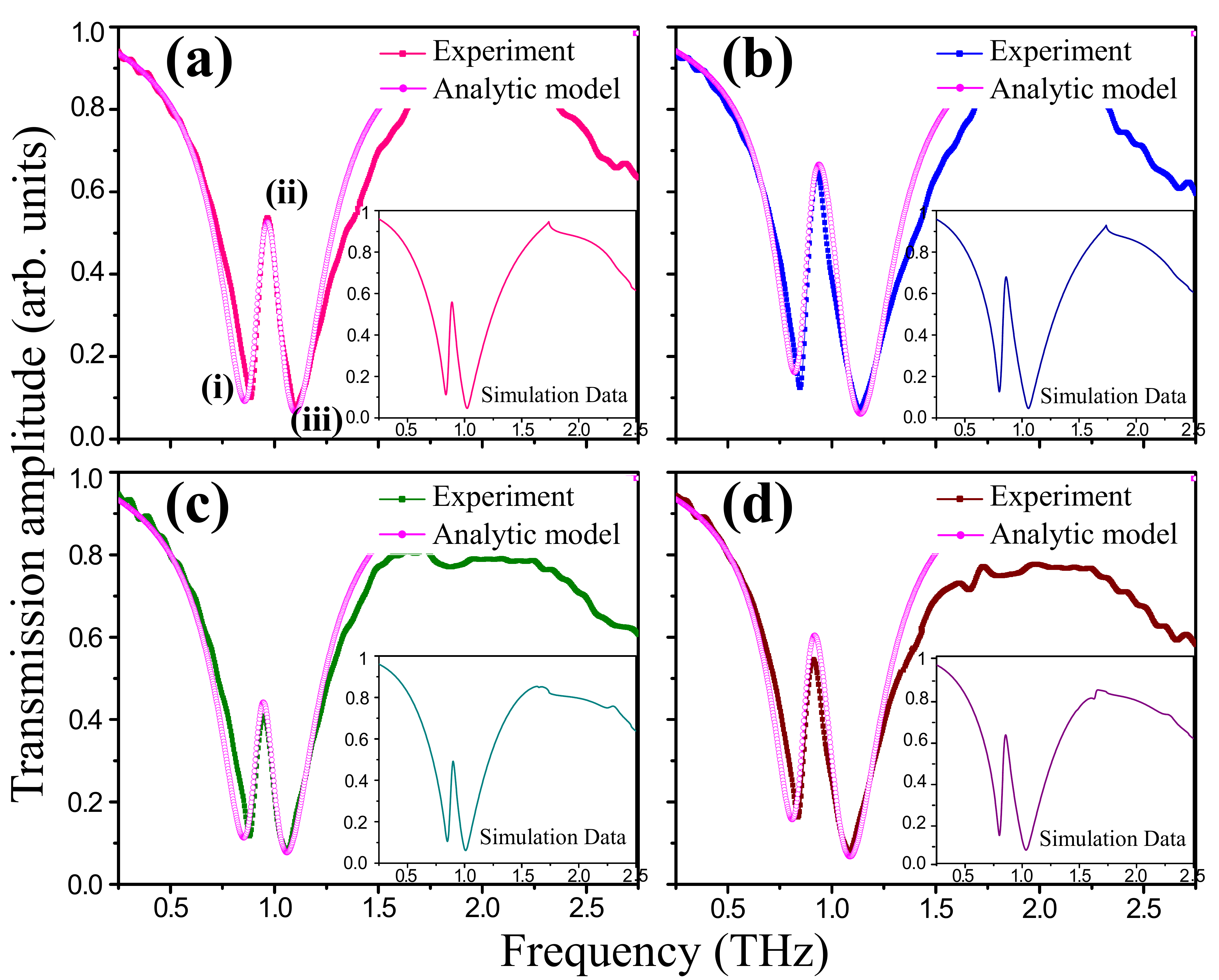}
\caption{\footnotesize{(color online). Terahertz transmission spectra for the incident $E_{x}$ field. Figures {\bf{(a)}}, {\bf{(b)}}, {\bf{(c)}} and {\bf{(d)}} show the experimentally measured as well as simulated transmission (inset) curves for metasurface samples MS-1, MS-2, MS-3 and MS-4 respectively.  Experimentally measured transmission curves are fitted with corresponding the analytic model data (colored magenta). ({\bf{i}}), ({\bf{ii}}) and ({\bf{iii}}) in ({\bf{a}}), simultaneously represents the antisymmetric, transmission and symmetric modes of the plasmonic hybridization, which is detailed in Fig.4(a).  } }
\end{center}
\end{figure}

\begin{figure*}[t]
  \begin{minipage}[c]{0.67\textwidth}
    \includegraphics[width=12cm]{Fig_4.pdf}
  \end{minipage}\hfill
  \begin{minipage}[c]{0.3\textwidth}
    \caption{
       \footnotesize{(color online).\textbf{(a)} Induced surface currents in SRR and the CRR at transmission dips ((i), 0.83THz  and (iii), 1.02THz) and at the transmission peak ((ii), 0.88THz) for the metasurface MS-1, are shown. \\
\textbf{(b)} $E_{x}$-field distributions at two transmission dips ((i) $\&$ (iii)) and at the transmission peak (ii) are shown for each metasurface samples. \\
\textbf{(c)} Depicts the surface current distributions for the metasurfaces MS-1, MS-2, MS-3 and MS-4 at their respective transmission peaks. }
    } \label{fig4}
  \end{minipage}
\end{figure*}

Fig.3(a) depicts EIT like transmission spectrum for MS-1, where, between two transmission dips (at 0.83THz and 1.02THz) a sharp transmission peak is observed at the frequency 0.088THz, which signifies the hybridization model of the plasmonic coupling via near fields of the individual resonators. The resonance dip at the lower frequency is the (i) antisymmetric mode and the higher frequency resonance dip is the (iii) symmetric mode. This behaviour is verified by simulating the surface current distribution as shown in the Fig.4(a) for MS-1, where at the transmission dips ((i) $\&$ (iii)), surface currents in the SRR and CRR run antiparallel and parallel to each other respectively for the (i) antisymmetric and the (iii) symmetric modes. Existence of the antisymmetric mode at the lower frequency signifies strong transverse dipole-dipole interaction within the coupled system. This dipole-dipole interaction can be electric as well as magnetic as the respective dipoles (electric/magnetic) are aligned in the transverse way to the mode separation line. Electric dipoles are aligned antiparellel to each other in the plane of the material and the induced magnetic dipoles are directed perpendicular to the metamaterial surface.

The observed transparency peaks (Fig.3) can be well explained using the induced surface currents and electric field distribution as shown in Fig.4(a $\&$ b). Fig.4(a) shows the the induced surface currents in the coupled system, where at the (ii) transparency peak induced circular currents in the SRR destructively interferes with the dipolar currents of CRR. As an evidence of the destructive interference, we observe extremely weak surface currents in the CRR that are opposite in direction to the surface current in the SRR. The same can be explained using electric field distributions (see Fig.4(b)) at the (i)antisymmetric resonance, (ii)transparency peak, and the (iii) symmetric resonance mode for each metasurface sample. By comparing the maximum electric field strength in the SRR gap at the (ii) transmission peaks with the transmission dips in (i) and (iii), clearly suggests that within the transparency peak SRR's {\emph{LC}} resonance dominates the dipolar CRR resonance. Thus the surface current and the electric field appears to be localized within the SRR of the coupled system. 

Introducing SRR-gap asymmetry in the system (MS-3) lead to a suppression in the transmission amplitude of the resulting transparency peak (Fig.3(c)). This suppression in the transparency peak is solely due to the weakened SRR resonance (ref blue curve in Fig.2(a)), because of its structural asymmetry. Displacing the gap to one end of the SRR arm results in rather a flaccid and weak electric field distribution (Fig.2(c)) at the SRR gap. This weakens the capacitive coupling within the SRR ring. As a result, the effective strength of the quasi-dark mode (SRR) decreases and results in a reduced transmission. Owing to the interference effects in EIT phenomenon, this result can be seen as analogous to the waves' interference, where decreasing the amplitude of one of the wave, results in the decreased strength of interference pattern. The same explanation holds true for the observed suppression in the transmission for MS-4 (Fig.3(d)) compared with the transmission of MS-2 (Fig.3(b)).

Upon introducing SRR-position asymmetry in the system, transmission through MS-2(4) shows an enhancement over the transmission observed for MS-1(3), as shown in Fig.3(b $\&$ d). Enhanced transmission is due to increased coupling between the CRR and the inner SRR, which results in strong destructive interference of the two fields at the transmission peak. This effect can be explained using Fig.4(c), which depicts the change in the strength of induced surface currents at the transparency peaks, for all the four coupled metasuface structures. For MS-1(3), strongly confined fields of the SRR induces opposing currents in CRR, which results in a destructive interference of the fields, giving rise to a sharp transmission peak. As the SRR ring is moved upwards (MS-2(4)), the separation between the bright (CRR) and the quasi-dark (SRR) modes decreases, which result in reduced current density in CRR. This is because, when the SRR gap (stronger E-field confinement) comes closer to the bright mode (top arm of the CRR), electric coupling between the two modes dominates the interaction. This reduced separation enhances the effective coupling in the system that leads to a strong cancellation of the opposing currents (enhanced destructive interference) within the coupled modes, resulting in an increased transmission of the incident field. This effect is also reflected in the {\emph{E}}-field distribution diagram as shown in Fig.4(b)(ii), where at the transparency peaks for the metasurfaces MS-2 and MS-4, the 
in the coupled system is decreased (compared with MS-1 and MS-2) as the result of the enhanced destructive interference between the two modes. Thus the entire system behaves as superadiative system (smaller {\emph{Q}}-value) due to the increased coupling within the coupled system. On the other hand, vice versa holds true, when the SRR ring is displaced downwards. The observed frequency red shift of the transmission peak and the asymmetric mode for the MS-2 and MS-4 (see Fig.3), indicates the increased electric field strength within the coupled metamaterial system\cite{40}. Thus by moving the quasi-dark mode relative to the bright mode, we can modulate the electromagnetically induced transparency by tailoring the strength of the electric and magnetic coupling within the system.

Effect of coupling on the transmission of light in the hybridized metamaterial system is theoretically studied by using two particle model. In the following coupled differential equations, we consider both particles (bright ($x_{b}$) and quasi-dark ($x_{d}$)) interact with the incoming electric field $E = E_{0} e^{i\omega t}$. 

\begin{equation}
\ddot{x}_{b}(t) + \gamma_{b} \dot{x}_{b}(t) + \omega^{2}_{b} x_{b}(t) + \Omega^{2}  x_{d}(t) = \frac{Q E}{M}
\end{equation}
\begin{equation}
\ddot{x}_{d}(t) + \gamma_{d} \dot{x}_{d}(t) + \omega^{2}_{d} x_{d}(t) + \Omega^{2}  x_{b}(t) = \frac{q_{d} E}{m_{d}}
\end{equation} 
 
Here, ($Q$, $q_{d}$), ($M$, $m_{d}$), ($\omega_{b}$, $\omega_{d}$) and ($\gamma_{b}$, $\gamma_{d}$)  are the effective charge, effective mass, resonance angular frequencies and the loss factors of the bright and the quasi-dark respectively. $\Omega$ defines the coupling strength between the bright and quasi-dark particles. In the above coupled equations, we substitute $q_{d}=\frac{Q}{A}$ and $m_{d}=\frac{M}{B}$, which gives the relative coupling of incoming radiation with the bright and the quasi-dark modes. Here, $A$ and $B$ are numerical constants. Now by expressing the displacements vectors for bright and quasi-dark modes as $x_{b}=c_{b} e^{i \omega t}$ and $x_{d}=c_{d} e^{i \omega t}$, we solve the above coupled equations (1) and (2) for $x_{b}$ and $x_{d}$.  

\begin{equation}
x_{b} = \frac{((B/A) \Omega^{2} + (\omega^{2} -  \omega^{2}_{d} + i \omega \gamma_{d}))(QE/M)}{\Omega^{4} - (\omega^{2} - \omega^{2}_{b} + i \omega \gamma_{b})(\omega^{2} - \omega^{2}_{d} + i \omega \gamma_{d})}
\end{equation}
and 
\begin{equation}
x_{d} = \frac{( \Omega^{2} + (B/A) (\omega^{2} - \omega^{2}_{b} + i \omega \gamma_{b}))(QE/M)}{\Omega^{4} - (\omega^{2} - \omega^{2}_{b} + i \omega \gamma_{b})(\omega^{2} - \omega^{2}_{d} + i \omega \gamma_{d})}
\end{equation}

The linear susceptibility ($\chi$), which relates the polarization (P) of the particle to the strength of incoming electric field (E) is expressed in terms of the displacement vectors as, $\chi = \frac{P}{\epsilon_{0} E} = \frac{Q x_{b} + q_{d} x_{d}}{\epsilon_{0} E}$.
\begin{equation}
\begin{split}
\chi = \chi_{R} + i\chi_{I} ~~~~~~~~~~~~~~~~~~~~ \\ \\
= \frac{K}{A^{2} B} ( \frac{A(B+1) \Omega^{2} + A^{2} (\omega^{2} - \omega^{2}_{d}) + B (\omega^{2} - \omega^{2}_{b})}{\Omega^{4} - (\omega^{2} - \omega^{2}_{b} + i \omega \gamma_{b})(\omega^{2} - \omega^{2}_{d} + i \omega \gamma_{d})} \\
+i\omega \frac{A^{2} \gamma_{d} + B \gamma_{b}}{\Omega^{4} - (\omega^{2} - \omega^{2}_{b} + i \omega \gamma_{b})(\omega^{2} - \omega^{2}_{d} + i \omega \gamma_{d})} )
\end{split}
\end{equation}

Here, Re[$\chi$] represents the dispersion and Im[$\chi$] gives the absorption (loss) within the medium. We fit 1-Im[$\chi$], to the experimental data shown in Fig.3 (colored magenta), which represents the transmission through a medium. For the fit, the values of the loss factors $\gamma_{b}$ and $\gamma_{d}$ are obtained from the linewidths of the curves shown in Fig.2(a), which are calculated to be around $3\times 10^{12}$ rad/sec and $5\times 10^{11}$ rad/sec respectively. The coupling strength $\Omega$ for each transmission curves is calculated using the formula given in Ref. \cite{33}, which can also be derived from Eqn.5 at the stop-band frequencies $\omega_{\pm}$, for a loss less medium (assuming $\omega_{b},\omega_{d} = \omega_{0}$). Using Eqn.5, at the transparency peak ($\omega=\omega_{T}$) where Re[$\chi$] = 0, we get , $\omega_{d} \approx \omega_{T}$ for larger $A$. At the stop-bands where Im[$\chi$]= $ \infty$, by using Eqn.5 we can arrive at the expression for $\omega_{b}$ (assuming $\omega_{d} = \omega_{T}$), $\omega_{b\pm}=\sqrt{\omega_{\pm} \mp \frac{\Omega^{4}}{\Omega^{2}_{\pm}}}$ . By substituting the calculated values for $\gamma_{b}$, $\gamma_{d}$, $\omega_{b}$, $\omega_{d}$, $\Omega$ and by putting $B = 2$ (mass of SRR is half the mass of CRR) and $K = 4\times 10^{25}$ (amplitude offset) in Eqn.5, we find an excellent fit for the transmission curves shown in Fig.3, for parameter $A = 40$. This suggests that, within these coupled metamaterial systems, interaction of the quasi-dark mode to the incoming radiation is nearly 20 times smaller than that of the bright mode. This seems reasonable considering an order of magnitude difference in their respective {\emph{Q}}-factors. Thus, susceptibility expression given in Eqn.5 provides close to experimental situations of the present system under consideration with very small deviations. To further evaluate coupling effects in these systems, we study the influence of coupling strength($\Omega$) on the {\emph{Q}}-factors of the transmission curves. We find that the {\emph{Q}}-factors obtained from the simulated transmission curves for the samples with different SRR positions (SRR-position asymmetries) follow $\frac{K}{\Omega^{2}}$ variation (Fig.5(a)), as predicted by the two particle model\cite{33}.

\begin{figure}
\begin{center}
  \includegraphics[width=8.5cm]{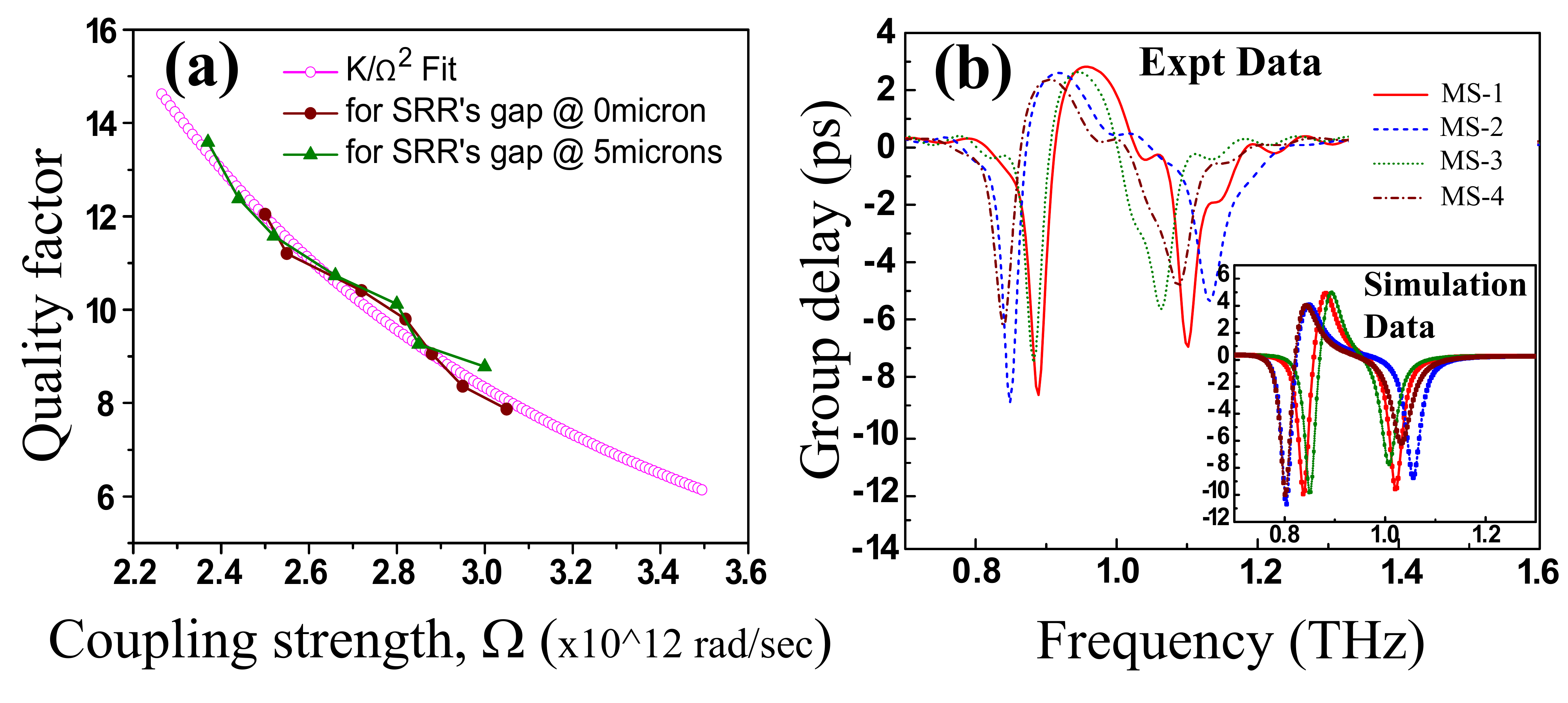}
\caption{\footnotesize{(color online). \textbf{(a)} K/$\Omega^{2}$ fit to the variation of Q-factors of the transmission curves with the coupling strength $\Omega$, under various SRR-position asymmetries of the metasurface samples. Brown curve represents Q-factor variation for metasurfaces with symmetric SRR ring, and green curve represents metasurfaces with asymmetric SRR ring. 
 \textbf{(b)} Both experimental and simulated group delay data for the incident radiation  within the transparency peak. }}
 \end{center}
\end{figure}

Figure 5(b) shows a variation of the measured group delay values for the transmission curves given in Fig.3. Experimentally measured values for group delay, delay band-width product (DBP) and the respective {\emph{Q}}-factors for the transmission curves for all the four metasurface samples are listed in Table.1. From the data we see that as coupling strength increases, DBP increases and the corresponding {\emph{Q}}-factor decreases and vice versa. For example, metasurface-2 that displays stronger mode coupling compared to all other samples, possesses maximum DBP and minimum {\emph{Q}}-factor. In addition to the tunability of the group delay, the proposed planar slow light metasurfaces offer improved DBP at the terahertz frequencies, indicating enhanced buffering capabilities for the broadband telecommunication networks. 

\begin{table}[h]
\begin{center}
\scalebox{1}{
\begin{tabular}{|c|c|c|c|} \hline
Samples &  Group delay  & DBP & Q-factor \\ 
 & $ t_{g} = \frac{-d\phi}{d\omega}$(ps) & $ ( t_{g} \times \Delta f )$  & \\ \hline
Sample-1 & 2.77 & 0.21 & 11.74 \\ \hline
Sample-2 & 2.52 & 0.217 & 9.83 \\ \hline
Sample-3 & 2.63  & 0.174 & 13.35 \\ \hline
Sample-4 & 2.35  & 0.18 & 11 \\ \hline
\end{tabular}}
  \caption{\footnotesize{Calculated group delay, DBP and Q-factor values for the experimentally measured transmission curves shown in Fig.3.}}
\end{center}
\end{table}

In summary, we have experimentally studied the near field effects on the transmission properties of planar metasurfaces under two types asymmetries in the system. We demonstrated that, introducing SRR-gap asymmetry suppresses the transmission, whereas in the system with SRR-position asymmetry, where the SRR gap is moved towards the bright mode, enhanced electric coupling will enhance the transmission. Our results show that the transmission amplitude can be modulated by varying the strength of electric and the magnetic dipoles. The effect of interplay between the strengths of electric and magnetic dipoles on the transmission provides a deeper insight into the classical analogy of the quantum interference effect. Using two particle model, we quantitatively analyzed the interactions within the system and the analytic data showed an excellent agreement with the observed results. The model also provides the relative coupling of the quasi-dark (SRR) mode to the incoming field in the near field coupled system. The proposed asymmetric planar slow light metasurfaces with tunable transparency characteristics will allow us to precisely control the group velocity of the pulse within the medium. They can be readily applied in broadband terahertz technologies and show potential applications as variable power attenuators, broadband filters, and compact delay lines for the terahertz waves.

\end{document}